\documentclass[10pt,twocolumn,amsmath,amssymb,aps,floatfix,preprintnumbers,nofootinbib,nobibnotes,bibnotes]{revtex4-2}
\usepackage{graphicx}% Include figure files
\usepackage{dcolumn}% Align table columns on decimal point
\usepackage{bm}% bold math
\usepackage{hyperref}% add hypertext capabilities
\usepackage[mathlines]{lineno}% Enable numbering of text and display math
\usepackage{epsfig}  
\usepackage{color}
\usepackage{slashed}
\usepackage{float}
\usepackage[table]{xcolor}

\large

\begin{document}

\title{Investigating the potential of $R_{K^{(*)}}^{\tau \mu}$ to probe lepton flavor universality violation}

\author{Ashutosh Kumar Alok}
\email{akalok@iitj.ac.in}
\affiliation{Indian Institute of Technology Jodhpur, Jodhpur 342037, India}

\author{Neetu Raj Singh Chundawat}
\email{chundawat.1@iitj.ac.in}
\affiliation{Indian Institute of Technology Jodhpur, Jodhpur 342037, India}

\author{Arindam Mandal}
\email{mandal.3@iitj.ac.in}
\affiliation{Indian Institute of Technology Jodhpur, Jodhpur 342037, India}

\begin{abstract}
In this work we study the potential of the lepton flavor ratios $R_{K}^{\tau \mu} \equiv \Gamma(B \to K \tau^+ \tau^-)/  \Gamma(B \to K \mu^+ \mu^-)$ and $ R_{K^{*}}^{\tau \mu} \equiv \Gamma(B \to K^* \tau^+ \tau^-)/  \Gamma(B \to K^* \mu^+ \mu^-)$ to probe lepton flavor universality (LFU) violation in $\tau-\mu$ sector. We show that these ratios can deviate from their SM values even if the new physics couplings are universal in nature, i.e., having equal couplings to $e$, $\mu$ and $\tau$ leptons. Therefore in order to utilize these observables to probe LFU violation, we need to compare the allowed range of $R_{K^{(*)}}^{\tau \mu}$ for class of solutions with only universal couplings to leptons and solutions having both universal and non-universal components. For the current $b \to s \ell \ell$ ($\ell=e,\,\mu$) data, we find that these two class of solutions can be discriminated provided the measured value of  $ R_{K^{*}}^{\tau \mu} $ is greater than the SM prediction. 
\end{abstract}
 
\maketitle 

\newpage

 %%%%%%%%%%%%%%%%%%%%%%%%%%%%%%%%%%%%
\section{Introduction} 
%%%%%%%%%%%%%%%%%%%%%%%%%%%%%%%%%%%%

The decays induced by the quark level transition $b \to s \ell \ell$ ($\ell=e,\, \mu, \,\tau $) have immense potential to probe physics beyond the Standard Model (SM) of electroweak interactions. This is due to multiple reasons. Firstly, within the SM, these decays can occur only at the loop level and hence have highly suppressed decay rates. Further, the same quark level transition induces a number of decay modes, such as $B \to K \ell \ell$, $B \to K^* \ell \ell$ and $B_s \to \phi \ell \ell$ decays. Therefore we are equipped with a plethora of observables to inspect new physics. Also, this decay mode is sensitive to $CP$ violation beyond the current paradigm \cite{Fleischer:2022klb,Fleischer:2023zeo} as within the SM the predicted values of various $CP$ violating observables are highly suppressed. Moreover, the decay channels $b \to s \ell \ell$ ($\ell=e,\, \mu$) have already started providing several enthralling hints of beyond SM physics.

The most striking deviation from the SM is revealed through the measurement of the branching ratio of  $B_s \to \phi\, \mu^+\,\mu^-$ decay. The measured value deviates from the SM prediction  at the level of  3.5$\sigma$  \cite{bsphilhc2,bsphilhc3}.  The decay $B \to K^* \, \mu^+\,\mu^-$ also  displays tension with the SM through the measurement of  the  optimized angular observable  $P'_5$ in  4.0 $\mathrm{GeV}^2 \le q^2 \le$ 6.0 $\mathrm{GeV}^2$ bin. The dissidence with the SM is at 3$\sigma$  level \cite{Kstarlhcb1,Kstarlhcb2,LHCb:2020lmf,sm-angular}. These anomalous measurements can be reconciled by assuming new physics in $b \to s \mu^+ \mu^-$ transition. 

 The lepton flavor ratio observables $R_{K} \equiv \Gamma(B^+ \to K^+ \mu^+ \mu^-)/\Gamma(B^+ \to K^+ e^+ e^-)$ and  $R_{K^*} \equiv \Gamma(B^0 \to K^{*0} \mu^+ \mu^-)/\Gamma(B^0 \to K^{*0} e^+ e^-)$ were defined to capture the mismatch  between $b \to s \mu^+ \mu^-$ and $b \to s e^+ e^-$ transitions. The measured values of these observables  relinquished tension with the SM \cite{LHCb:2021trn,rkstar}. The explanation of data required new physics couplings to be non-universal in nature, i.e., the new physics couplings in electron and muon sectors should be different. The favoured Lorentz structures of such new physics in $b \to s \ell \ell$ ($\ell=e,\, \mu $)  were determined through  a model independent global analysis of $b \to s \ell \ell$  data  using the language of effective field theory \cite{Descotes-Genon:2013wba,Altmannshofer:2013foa,Hurth:2013ssa,Hiller:2014yaa,Capdevila:2016ivx,Ciuchini:2017mik,Alok:2017jgr,Alok:2019ufo,Altmannshofer:2021qrr,Datta:2019zca,Carvunis:2021jga,Alguero:2021anc,Geng:2021nhg,Hurth:2021nsi,Angelescu:2021lln,Alok:2022pjb,Gangal:2022ole,Ciuchini:2021smi}.  In most of the analysis the non-universality was accomplished by assuming new physics only in the muon sector whereas some analyses assumed non-equal couplings to both muons and electrons. In \cite{Alguero:2018nvb}, it was shown that apart from non-universal couplings (only in muon sector), one can also have universal couplings, i.e., new physics which couples equally to electron, muon and tauon. In fact such class of new physics scenarios provided a better fit as compared to scenarios having only non-universal component \cite{Alguero:2018nvb,Alguero:2019ptt,Alguero:2021anc,Alguero:2022wkd}. 

However, the recent updates from the LHCb collaboration in December 2022 \cite{LHCb:2022qnv,LHCb:2022zom} accoutered  values of $R_{K}$ and $R_{K^*}$ which now concur with the SM prediction. This forces the new physics couplings to be nearly universal in nature \cite{SinghChundawat:2022ldm,SinghChundawat:2022zdf,Ciuchini:2022wbq,Alguero:2023jeh,Wen:2023pfq,Allanach:2023uxz,Li:2023mrw}. If couplings are universal  then they will also generate new physics effects in  $b \to s \tau^+ \tau^-$ decay channel. A legitimate question to ask at this stage is whether the current data in $b \to s e^+ e^-$ and $b \to s \mu^+ \mu^-$ sectors can allow for lepton flavor universality (LFU) violation in the $\tau-\mu$ sector. The formalism developed in \cite{Alguero:2018nvb} is equipped to allow such an inquisition as it has provision for both universal as well as non-universal components. 

In the current work we examine the new physics potential of lepton flavor ratio observables $R_{K}^{\tau \mu} \equiv \Gamma(B \to K \tau^+ \tau^-)/  \Gamma(B \to K \mu^+ \mu^-)$ and $ R_{K^{*}}^{\tau \mu} \equiv \Gamma(B \to K^* \tau^+ \tau^-)/  \Gamma(B \to K^* \mu^+ \mu^-)$.  In particular, we inspect how well these observables can discriminate between the class of solutions having only universal couplings and solutions having both universal  and non-universal new physics couplings.    Owing to the nomenclature, the  LFU ratios  are expected to render values within their predicted SM range for  new physics with universal couplings. The ratios  $R_{K}$ and $R_{K^*}$ are the most popular examples of such LFU ratios \cite{Hiller:2003js,Bordone:2016gaq,Isidori:2020acz,Isidori:2022bzw,Nabeebaccus:2022pje}. Therefore  it is natural to expect the same for the ratio observables $R_{K}^{\tau \mu} $ and $ R_{K^{*}}^{\tau \mu} $. 

In this work we show that ratios $R_{K}^{\tau \mu} $ and $ R_{K^{*}}^{\tau \mu} $ can engender values beyond their predicted SM values {\it even for} new physics solutions with  universal couplings. Therefore in order to discriminate between the class of solutions having only universal couplings and solutions having both universal and non-universal components, we need to compare the allowed range for these class of solutions. If the allowed range is distinct for the two classes,  only then $R_{K^{(*)}}^{\tau \mu}$ can serve the purpose of discriminating between the LFU conserving and violating new physics. We explore this discriminating ability of $R_{K}^{\tau \mu} $ and $ R_{K^{*}}^{\tau \mu} $ by making use of current experimental measurements in $b \to s e^+ e^-$ and $b \to s \mu^+ \mu^-$ sectors. 

The plan of work is as follows. In the next section, we discuss the framework of universal and non-universal new physics in $b \to s \ell \ell$  decay. We also provide fit results, i.e we provide allowed parameter space for new physics couplings for class of solutions with only universal component as well as for class of solutions having both universal as well as non-universal components. In Sec.~\ref{results}, we  discuss $R_{K}^{\tau \mu} $ and $ R_{K^{*}}^{\tau \mu} $ results for the two classes of new physics. The conclusions are presented in Sec.~\ref{concl}.

 %%%%%%%%%%%%%%%%%
 
\section{Formalism and Fit Results}

The effective Hamiltonian for $ b\to s \ell^+  \ell^- $ ($\ell=e,\, \mu, \,\tau $) transition within the SM can be written as
\begin{eqnarray}
\mathcal{H}^{\rm SM}_{\rm eff} &=& - \frac{\alpha_{em} G_F}{\sqrt{2} \pi} V_{ts}^* V_{tb}  \nonumber\\ && \times\Big[2 \frac{C_7^{\rm eff}}{q^2}
 [\overline{s} \sigma^{\mu \nu} q_\nu (m_s P_L  + m_b P_R)b ] \bar{\ell} \gamma_\mu \ell
\nonumber\\ && + 
C_9^{\rm eff} (\overline{s} \gamma^{\mu} P_L b)(\overline{\ell} \gamma_{\mu} \ell)  + C_{10} (\overline{s} \gamma^{\mu} P_L b)(\overline{\ell} \gamma_{\mu} \gamma_5 \ell) 
\Big]\nonumber\\ &&  + H.c. \,.
\end{eqnarray}
Here $P_{L,R} = (1 \mp \gamma_{5})/2$ and  $q$ in the first term is the momentum of the off shell photon in the effective $b \to s \gamma^*$ decay. Further, $V_{ts}$ and $V_{tb}$ are the elements of the quark mixing matrix, $\alpha_{em}$ is the fine-structure constant and  $G_F$ is the Fermi constant.  

We now assume new physics in the form of vector  and axial-vector for which the effective Hamiltonian for $b \to s \ell^+ \ell^-$ decay can be written as 
\begin{eqnarray}
\mathcal{H}^{\rm NP}_{\rm eff} &=& -\frac{\alpha_{\rm em} G_F}{\sqrt{2} \pi} V_{ts}^* V_{tb} \left[ C_{9\ell} (\overline{s} \gamma^{\mu} P_L b)(\overline{\ell} \gamma_{\mu} \ell) \right. \nonumber \\
& & \left. + C_{10\ell} (\overline{s} \gamma^{\mu} P_L b)(\overline{\ell} \gamma_{\mu} \gamma_5 \ell)  + C^{\prime}_{9\ell} (\overline{s} \gamma^{\mu} P_R b)(\overline{\ell} \gamma_{\mu} \ell)\right. \nonumber \\
& & \left. + C^{\prime}_{10\ell} (\overline{s} \gamma^{\mu} P_R b)(\overline{\ell} \gamma_{\mu} \gamma_5 \ell)\right] + H.c.  \,\,,
\label{HNP}
\end{eqnarray} 
where $C_{(9,10)\ell}$ and $C^{\prime}_{(9,10)\ell}$ are the new physics WCs having both universal and non-universal components: 
\begin{eqnarray}
C_{(9,10)e}&=&C_{(9,10)\tau}=C_{(9,10)}^U\,,\nonumber\\
C_{(9,10)e}^{\prime}&=&C_{(9,10)\tau}^{\prime}=C_{(9,10)}^{\prime U}\,,\nonumber\\
C_{(9,10)\mu}&=& C_{(9,10)}^U +C_{(9,10)\mu}^V\,,\nonumber\\
C_{(9,10)\mu}^{\prime}&=& C_{(9,10)}^{\prime U} +C_{(9,10)\mu}^{\prime V}\,. 
\end{eqnarray}
Here $C_{(9,10)}^U$ and $C_{(9,10)}^{\prime U}$ are the universal contributions to the WCs. These contribute equally to all  $b \to s \ell^+ \ell^-$ transitions whereas $C_{(9,10)\mu}^V$ and $C_{(9,10)\mu}^{\prime V}$ can contribute only to $b \to s \mu^+ \mu^-$ decay. Therefore there can be three possibilities: 
\begin{itemize}
\item $C_{(9,10)}^U=C_{(9,10)}^{\prime U}=0$, i.e we only have non-universal couplings. This scenario is disfavoured by the current data, in particular the updated measurements of $R_K$ and $R_{K^*}$ by the LHCb collaboration which is now consistent with their SM predictions.

\item $C_{(9,10)\mu}^V=C_{(9,10)\mu}^{\prime V}=0$, i.e we only have universal couplings. We call this as framework-I (F-I).

\item both universal as well as non-universal couplings are present. We call this as framework-II (F-II).

\end{itemize}

\begin{table}[hbt]
\addtolength{\tabcolsep}{-1pt}
\begin{center}
\begin{tabular}{|c|c|c|c|}
  \hline\hline
F-I Solutions	 & WCs & 1$\sigma$ range & $\Delta \chi^2$  \\
  \hline
SU-I  & $C^{U}_9$ & $-1.08 \pm 0.18$ & 27.90 \\ 
\hline
SU-II & $C^{U}_9 = - C^{U}_{10}$ & $-0.50 \pm 0.12 $ &  18.85 \\ 
\hline
 SU-III  & $C^{U}_9 = - C^{'U}_{9}$ &$-0.88 \pm 0.16$ & 26.92 \\
 \hline
\end{tabular}
\caption{Allowed new physics solutions assuming new physics couplings to be universal. Here $\Delta\chi^2 = \chi^2_{\rm SM}-\chi^2_{\rm bf}$ where $\chi^2_{\rm bf}$ is the $\chi^2$ at the best fit point and $\chi^2_{\rm SM}$  corresponds to the SM which is $\chi^2_{\rm SM}  \approx$ 184.} 
\label{scenarios1}
\end{center}
\end{table}

Within framework-I, assuming contributions from one operator or two related operators at a time, the scenarios favored by the current data along with the 1$\sigma$ range of the WCs, as obtained in \cite{SinghChundawat:2022zdf}, are listed in Table.~\ref{scenarios1}. The parameter space of the WCs are determined by performing a global fit to  179 observables in $b \to s \ell^+ \ell^-$ decay.  These include the updated measurements of $R_K$ and $R_{K^*}$ by the LHCb Collaboration in December, 2022 \cite{LHCb:2022qnv,LHCb:2022zom}  along with a number of $CP$ conserving $b \to s \mu^+ \mu^-$ and $b \to s e^+ e^-$ observables.  The fit also includes  the new world average of the branching ratio of $B_s \to \mu^+ \mu^-$ which is $(3.45 \pm 0.29)\times 10^{-9}$ \cite{Ciuchini:2022wbq}. This resulted due to the recently updated  measurement  by the CMS collaboration using the full  Run 2 dataset \cite{CMS:2022mgd}. The updated world average of the branching ratio of $B_s \to \mu^+ \mu^-$ is now  in agreement with its SM prediction \cite{Bobeth:2013uxa,UTfit:2022hsi}. The complete list of observables used in the fit along with the fitting methodology is provided in \cite{SinghChundawat:2022ldm}.

\begin{table}[hbt]
\addtolength{\tabcolsep}{-1pt}
\begin{center}
\begin{tabular}{|c|c|c|c|}
  \hline\hline
F-II Solutions	 & WCs & 1$\sigma$ range  & $\Delta \chi^2$ \\
  \hline
S-V  & $C^{V}_{9\mu}$ & (-1.31, -0.53 )  &  \\ 
     & $C^{V}_{10\mu}$ & (-0.66 ,0.07)& 20.25\\
     & $C^{U}_9 = C^{U}_{10}$ &  (-0.13, 0.58) &  \\
\hline
S-VI & $C^{V}_{9\mu} = - C^{V}_{10\mu}$ & (-0.33, -0.20)   &\\ 
     & $C^{U}_9 = C^{U}_{10}$ & (-0.43, -0.17) & 16.81 \\
 \hline
S-VII  & $C^{V}_{9\mu}$ & (-0.43,  -0.08) & \\ 
     & $C^{U}_9$ & (-1.07, -0.58) & 30.25 \\
\hline
S-VIII  & $C^{V}_{9\mu} = - C^{V}_{10\mu}$  & (-0.18,  -0.05) &  \\ 
     & $C^{U}_9$ & (-1.15,-0.77) & 31.36 \\
\hline
\hline
S-IX  & $C^{V}_{9\mu} = - C^{V}_{10\mu}$ &(-0.27,-0.12) & \\ 
     & $C^{U}_{10}$  & (-0.09,0.27) & 12.96    \\
\hline
S-X  & $C^{V}_{9\mu}$ &(-0.72,-0.41)& \\ 
     & $C^{U}_{10}$   &(0.05,0.34)& 21.16 \\
\hline
S-XI  & $C^{V}_{9\mu}$ & (-0.82, -0.51)&  \\ 
     & $C^{\prime U}_{10}$  & (-0.26,-0.04)  & 21.16  \\
\hline
S-XIII  & $C^{V}_{9\mu}$ & (-0.96,-0.60) &  \\ 
      & $C^{\prime V}_{9\mu}$ &(0.22,0.63) &  \\ 
   & $C^{ U}_{10}$ &(0.01,0.38) &\\ 
   & $C^{\prime U}_{10}$ &(-0.08,0.24) & 26.01 \\ 
 \hline
\end{tabular}
\caption{Allowed new physics solutions assuming both universal as well as non-universal new physics couplings. Here $\Delta\chi^2 = \chi^2_{\rm SM}-\chi^2_{\rm bf}$ where $\chi^2_{\rm bf}$ is the $\chi^2$ at the best fit point and $\chi^2_{\rm SM}$  corresponds to the SM which is $ \approx$ 184.}
\label{scenarios2}
\end{center}
\end{table}

\begin{figure*} [htb]
\centering
\includegraphics[scale=.60]{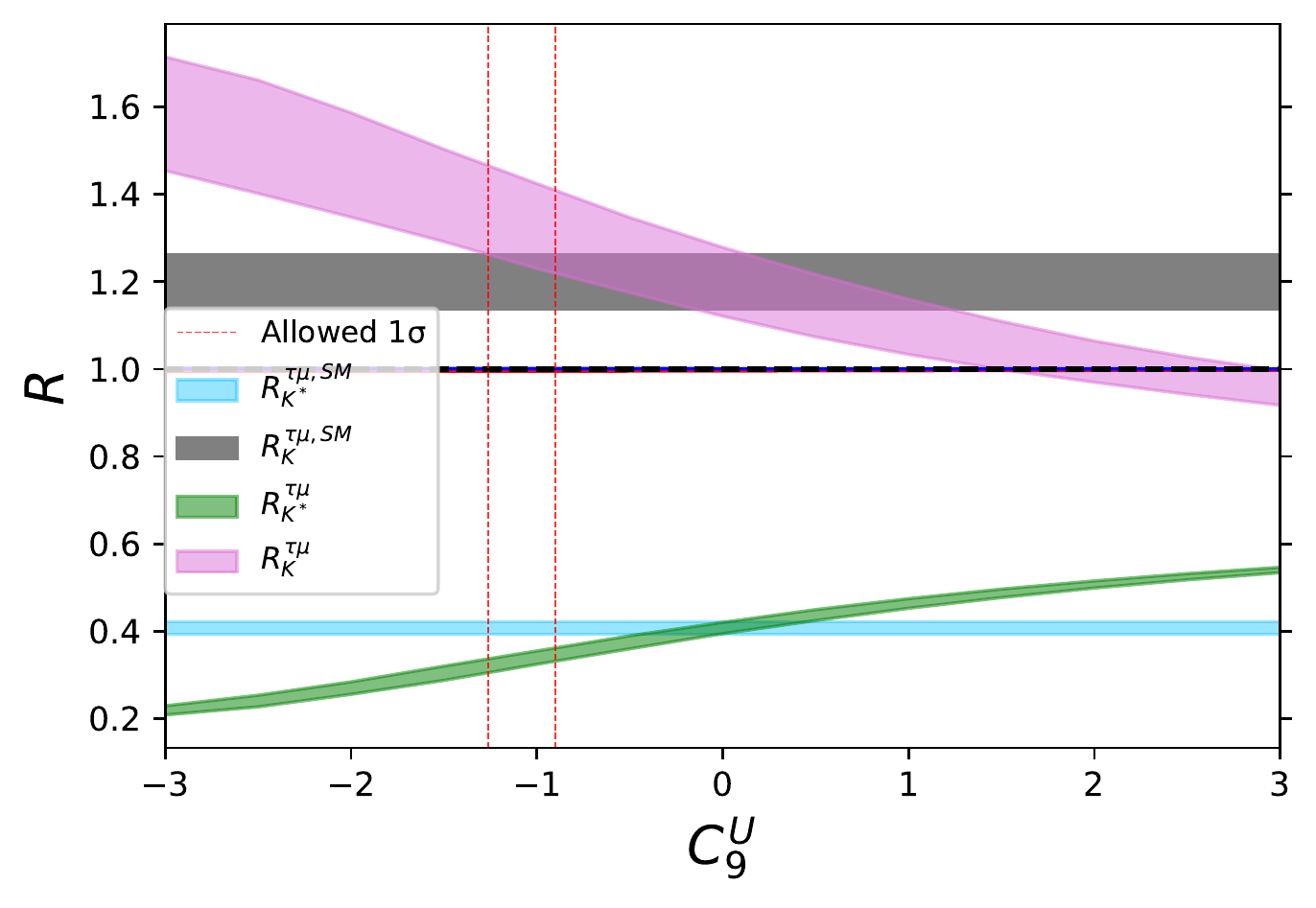}
 \includegraphics[scale=.60]{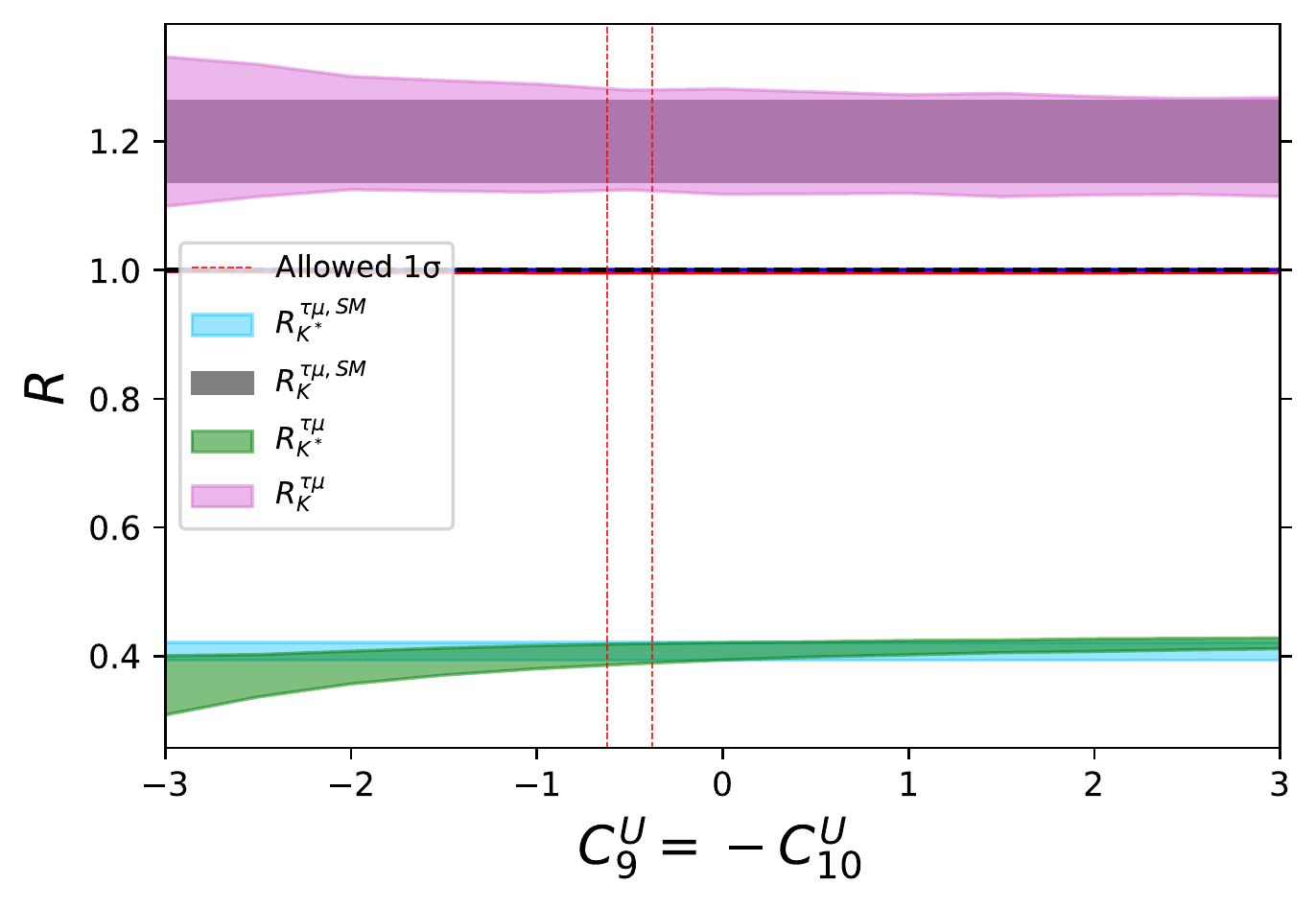}\\
\includegraphics[scale=.60]{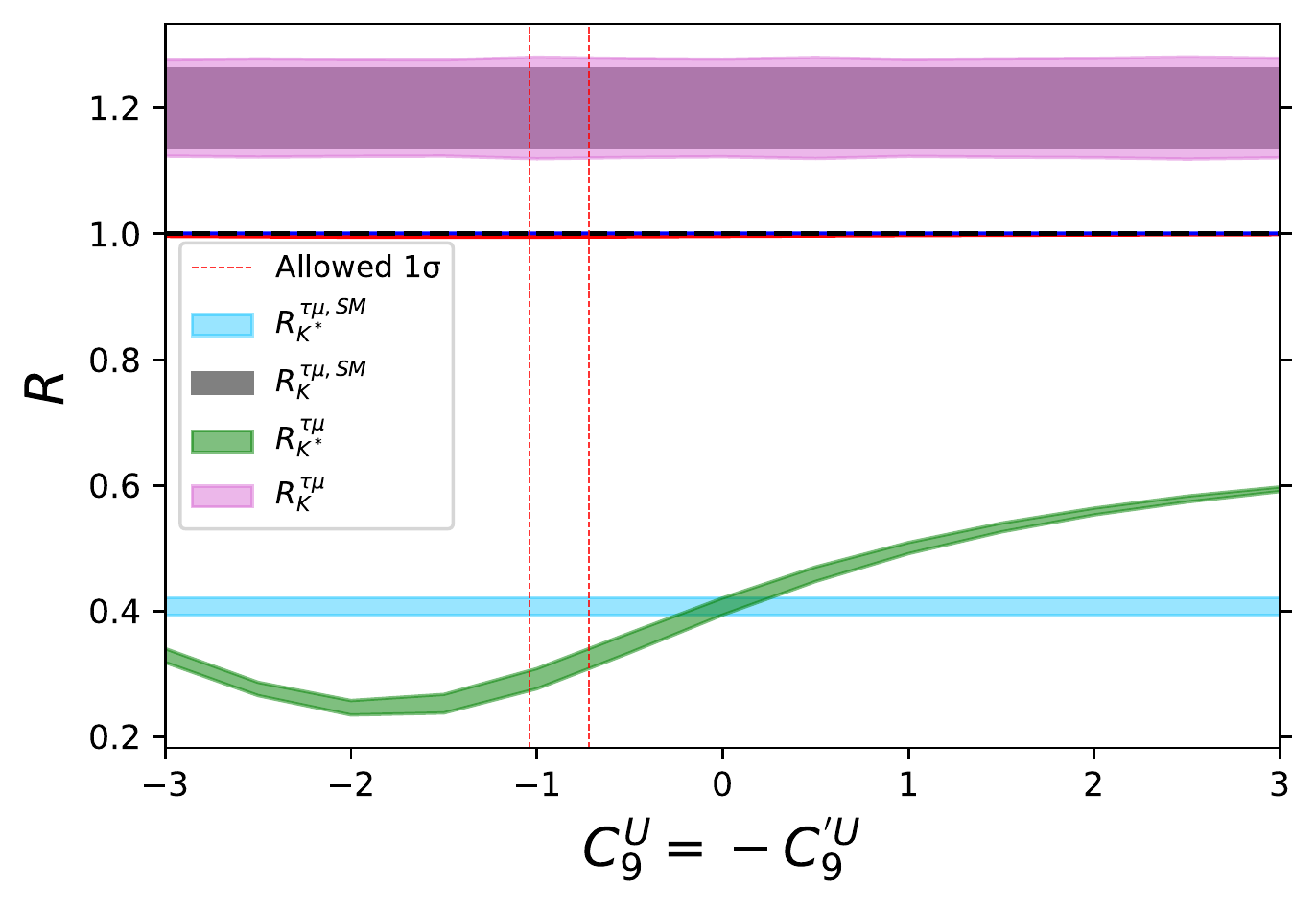}
\caption{The figure delineates the functional dependence of  $R_{K}^{\tau \mu} $ and $ R_{K^{*}}^{\tau \mu} $ ratios on the new physics WCs having only universal components. The top-left, top-right and the bottom panel corresponds to the $C_9^U$,  $C_9^U = - C_{10}^U$  and  $C_9^U = - C_{9}^{'U}$ new physics solutions, respectively. The grey (light magenta) and blue (light green) band correspond to the SM (new physics) predictions for  $R_{K}^{\tau \mu} $ and $ R_{K^{*}}^{\tau \mu} $, respectively. The band is due to the theoretical uncertainties. The vertical dashed lines are the 1$\sigma$ region allowed by the current experimental measurements in 
$b \to s \ell \ell$ ($\ell=e,\,\mu$) sectors. To illustrate the fact that the predicted values of $R_{K}^{ \mu e} \equiv R_K $ and $ R_{K^{*}}^{\mu e} \equiv R_{K^*} $ ratios are consistent with their SM predictions for universal couplings, we also show them in the above plots. The horizontal line at $R=1$ shows the SM predictions of $R_K$ and $R_{K^{*}}$. The new physics predictions for $R_K$ and $R_{K^{*}}$ are illustrated by red and blue bands, respectively which coincide with their SM predictions.    }
\label{fig:tau-mu-1}
\end{figure*}

For framework-II, a complete set of favored scenarios were identified in \cite{Alguero:2018nvb,Alguero:2019ptt,Alguero:2021anc}. These new physics solutions along with the updated 1$\sigma$ range of the WCs as obtained in \cite{SinghChundawat:2022ldm} are listed in Table \ref{scenarios2}. Here again the fit was performed using the same 179 observables which were used in the fit for F-I. The F-II solutions can be classified into two categories: Class-A and Class-B solutions. The class-A solutions are characterized by $C_9^U$ contributions and it has  four favored scenarios. Following the nomenclature of \cite{Alguero:2021anc}, these solutions are denoted as S-V, S-VI, S-VII and S-VIII.  The class-B scenarios are characterized either by $C_{10}^U$ or $C_{10}^{\prime U}$  contributions. The favoured solutions under this class are denoted as S-IX, S-X, S-XI and S-XIII. These scenarios can arise naturally in a number of new physics models, see for e.g. \cite{Crivellin:2019dun,Bobeth:2016llm,Crivellin:2018yvo}.

In the next section, we analyze the potential of $R_{K}^{\tau \mu} $ and $ R_{K^{*}}^{\tau \mu} $ ratios to probe LFU violation in the $\tau-\mu$ sector.

%%%%%%%%%%%%%%%%%
 \section{Results and Discussions}
\label{results}

The SM prediction for $R_{K}^{\tau \mu} $ observable for the $B^0 \to K^0$ decay mode in [15-22] $q^2$ bin is  \cite{Bouchard:2013mia,Du:2015tda,Belle-II:2018jsg,Straub:2018kue}
\begin{equation}
R_{K}^{\tau \mu,\, \rm SM} = 1.20 \pm 0.07 \,.
\end{equation}
 For $ R_{K^{*}}^{\tau \mu} $ observable, the SM prediction in [15 - 19] $q^2$ bin  for the $B^0 \to K^{*0}$ decay mode is \cite{Straub:2018kue}
 \begin{equation}
 R_{K^{*}}^{\tau \mu,\, \rm SM}= 0.41 \pm 0.01\,.
\end{equation}  
Like $R_{K}^{ \mu e} \equiv R_K $ and $ R_{K^{*}}^{\mu e} \equiv R_{K^*} $, these observables are expected to capture the possible mismatch between the $\tau-\mu$ sector. Therefore, naively speaking, one should expect   $R_{K^{(*)}}^{\tau \mu}$   observables  to render values within their predicted SM range for new physics solutions having 
only universal component. However, as we will show below, this is true only for a narrow region where the new physics WCs are close to zero, i.e. closer to the SM. 

\begin{figure*} [htb]
\centering
\includegraphics[scale=.9]{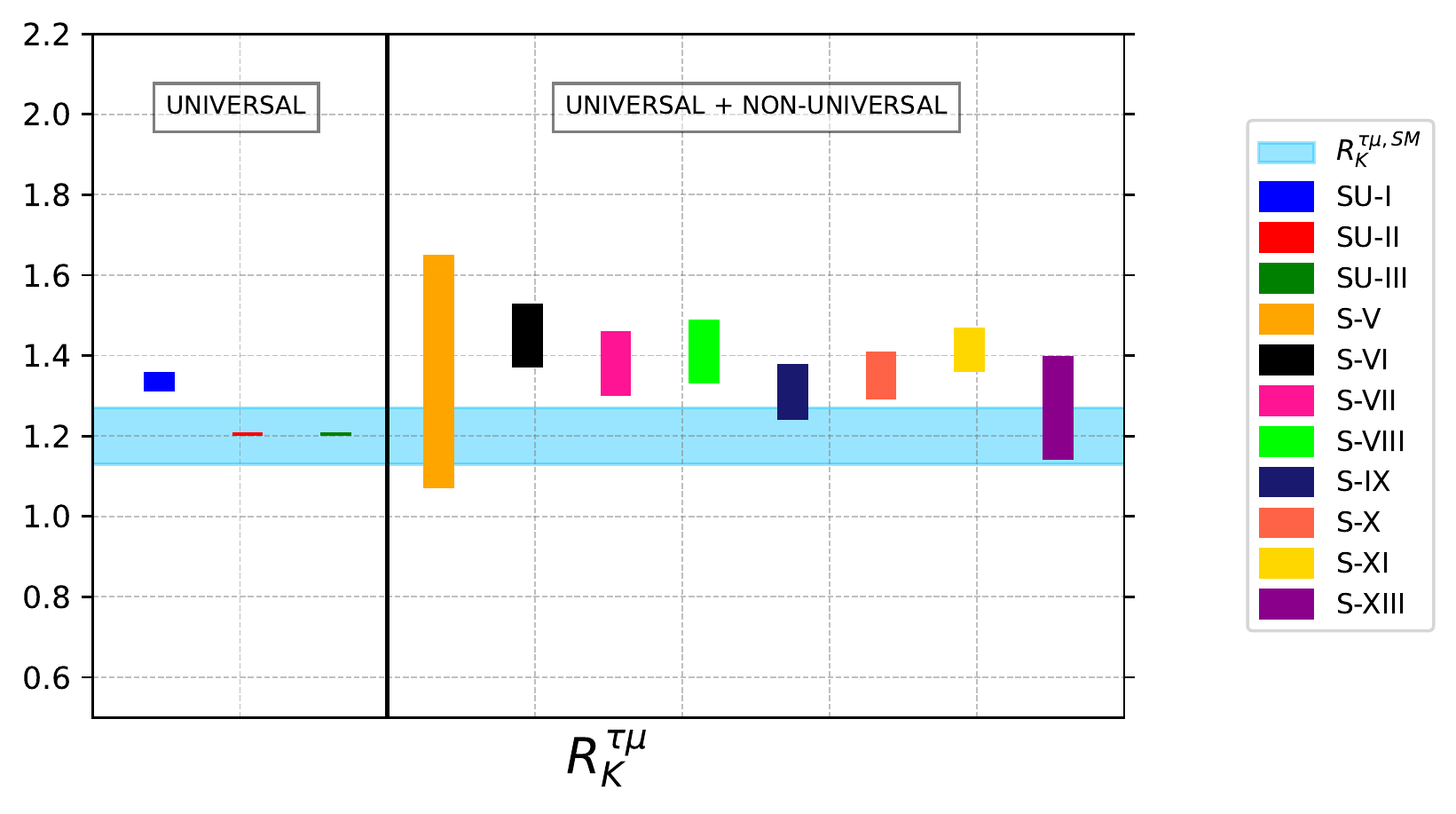}
\includegraphics[scale=.9]{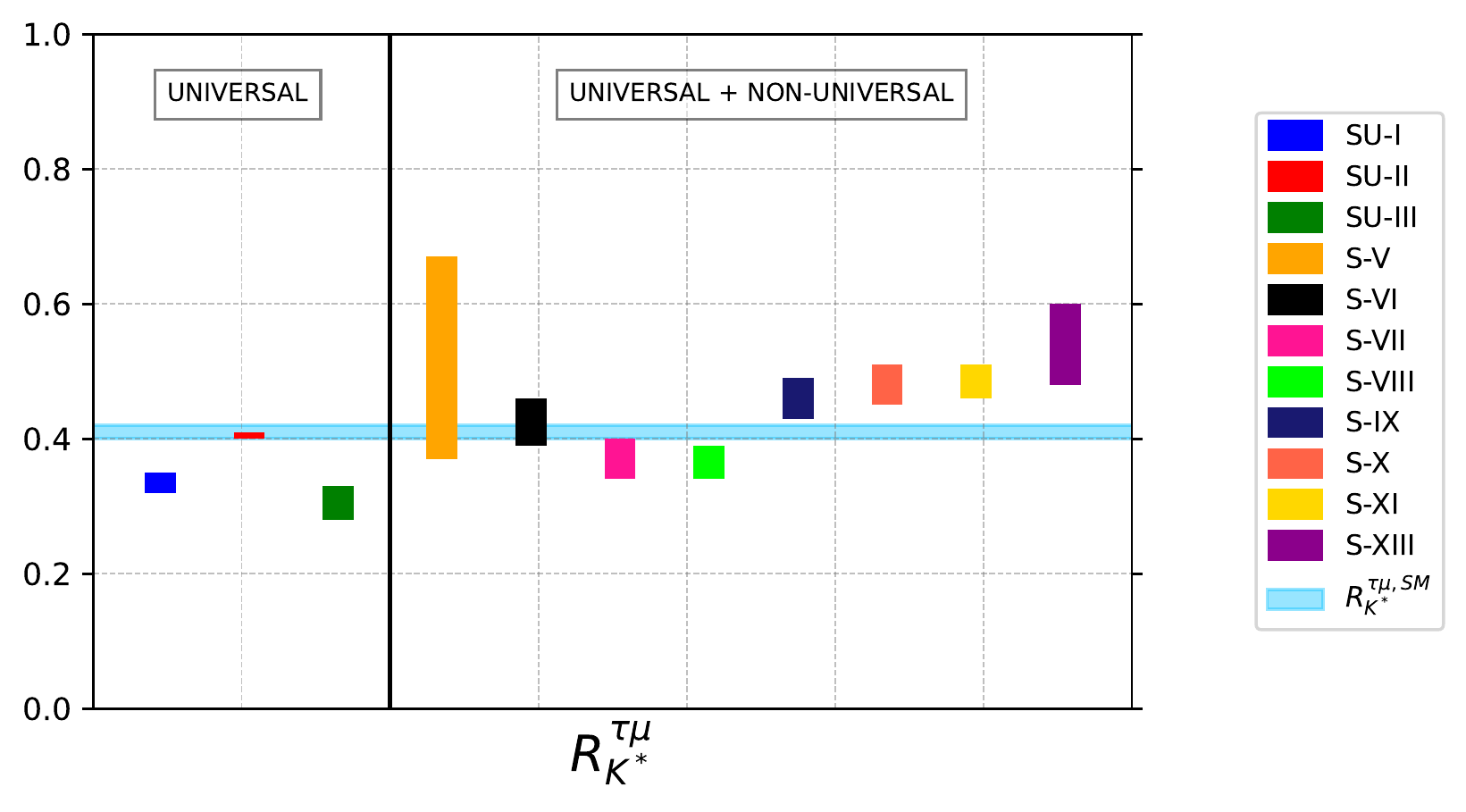}
\caption{Prediction of $R_{K}^{\tau \mu} $  and $ R_{K^{*}}^{\tau \mu} $ (1$\sigma$ range) for all allowed solutions corresponding to F-I (only universal couplings) and F-II (both universal and non-universal couplings) frameworks. These ratios are obtained using {\tt flavio} \cite{Straub:2018kue} where the observables are preimplemented based on refs. \cite{Bharucha:2015bzk,Gubernari:2018wyi}. }
\label{fig:tau-mu-2}
\end{figure*}

Fig.~\ref{fig:tau-mu-1} illustrates the functional dependence of these $R^{\tau \mu} $ ratios on the new physics WCs having only universal components. The three scenarios depicted in Fig.~\ref{fig:tau-mu-1} are favored by the current $b \to s \ell \ell$ ($\ell=e,\, \mu $) data. However, in order to understand the generic behaviour of these observables, we show regions of WCs much beyond what is allowed by the current data. 

It is apparent from the top-left panel of Fig.~\ref{fig:tau-mu-1} that $R_{K}^{\tau \mu}$ deviates from the SM in the entire range of considered WC except in a narrow range around $C_9^U \approx 0$. The deviation becomes large for larger values of $C_9^U$. Further, for region of $C_9^U$ favored by the current data, $R_{K}^{\tau \mu} > R_{K}^{\tau \mu,\,{\rm SM}}$ indicating that merely $R_{K}^{\tau \mu,\,{\rm NP}} \neq R_{K}^{\tau \mu,\,{\rm SM}}$ is not sufficient enough to capture LFU violation in  the $\tau-\mu$ sector. A similar feature is observed for the $ R_{K^{*}}^{\tau \mu} $ observable. Here also $R_{K^*}^{\tau \mu} \neq R_{K^*}^{\tau \mu,\,{\rm SM}}$ in the entire region under consideration except around $C_9^U \approx 0$. However, in this case $R_{K^*}^{\tau \mu} < R_{K^*}^{\tau \mu,\,{\rm SM}}$ in the region of WC allowed by the current data. For comparison, we also show predictions for $R_{K}$ and $R_{K^*}$. As expected, $R_{K}\approx R_{K^*} \approx 1$ in the entire region.

For $C_9^U = - C_{10}^U$ solution,  $R_{K}^{\tau \mu} \approx R_{K}^{\tau \mu,\,{\rm SM}}$ in the entire range of WCs under consideration. This also includes the 1$\sigma$ allowed region of $C_9^U = - C_{10}^U$. However for  $R_{K^*}^{\tau \mu}$, there is moderate deviation for values of WC less than $\approx$ -1. In the 1$\sigma$ allowed range, $R_{K^*}^{\tau \mu}$ is consistent with the SM prediction. Here again, $R_{K}$ and $R_{K^*}$ are consistent with the SM in the entire  range of WCs. 

The bottom panel of Fig.~\ref{fig:tau-mu-1} exemplifies functional dependence of $R_{K}^{\tau \mu}$ and $R_{K^*}^{\tau \mu}$ observables for the  $C_9^U = - C_{9}^{'U}$ scenario. As apparent from the plot, this scenario predicts $R_{K}^{\tau \mu}$ similar to the SM in the entire range of WCs under consideration. However, $R_{K^*}^{\tau \mu}$ observable show pronounced deviation from the SM prediction in the entire region barring a narrow range of WCs near zero. As expected, the $\mu-e$ ratios $R_{K}$ and $R_{K^*}$ are consistent with the SM in the entire range. 

The deviation of $R^{\tau \mu}$ from SM even for universal couplings can be attributed to {\it mass effects}. In order to understand this, we consider  $R_{K^{(*)}}^{\tau \mu}$, the ratio associated with $B \to {K^{(*)}}\ell^+\ell^-$ decay mode. The decay width, $\Gamma_{l}^{K} = \Gamma(B \to K\ell^+\ell^-)$, for $B \to K\ell^+\ell^-$ decay  in [15-22] $q^2$ bin  can be written as,
\begin{eqnarray}
\Gamma_{l}^{K}=A_{0}^{l}+  A_{1}^{l}  (C_{9}^{U} +C_{9}^{'U}) + A_{2}^{l} (C_{9}^{U} +C_{9}^{'U})^{2}  \nonumber \\ + A_{3}^{l} (C_{10}^{U} +C_{10}^{'U})  +A_{4}^{l} (C_{10}^{U} +C_{10}^{'U})^2.
\label{gamma}
\end{eqnarray}

The factors $A_{i}^{l}$'s (where $ i=0,1,2,3,4$) are primarily functions of form-factors and lepton masses \cite{Bobeth:2007dw,Becirevic:2012fy}.
 The approximate numerical values of $A_{i}^{l}$'s ($\times 10^{20}$) are obtained to be
\begin{eqnarray}
A_{i}^{\tau} \approx (4.11, 0.62, 0.08, -1.40, 0.16), \nonumber\\
A_{i}^{\mu} \approx (3.50, 0.80, 0.11, -0.92, 0.11),\nonumber\\
A_{i}^{e} \approx (3.49, 0.80, 0.11, -0.92, 0.11). 
\label{A}
\end{eqnarray}

 Similarly, the decay width of $B\to K^{*}\ell^{+}\ell^{-}$ in [15-19] $q^2$ bin can be written, in general, in terms of new physics WCs as, 
\begin{eqnarray}
\Gamma_{l}^{K^*}=B_{0}^{l}+  B_{1}^{l}  (C_{9}^{U} +C_{9}^{'U}) + B_{2}^{l} (C_{9}^{U} +C_{9}^{'U})^{2} \nonumber \\  + B_{3}^{l} (C_{10}^{U} +C_{10}^{'U})  +B_{4}^{l} (C_{10}^{U} +C_{10}^{'U})^2 \nonumber \\ + B_{5}^{l}  (C_{9}^{U} - C_{9}^{'U}) + B_{6}^{l} (C_{9}^{U}-C_{9}^{'U})^{2} \nonumber \\ + B_{7}^{l} (C_{10}^{U} -C_{10}^{'U})  +B_{8}^{l} (C_{10}^{U} -C_{10}^{'U})^2.
\label{gamma1}
\end{eqnarray}
The values of the functions $B$'s are listed in table \ref{Bvalues}.

{\rowcolors{2}{white!65!white!30}{gray!20!white!30}
\begin{table*}[hbt]
\addtolength{\tabcolsep}{-1pt}
\begin{center}
\begin{tabular}{|c||c|c|c|c|c|c|c|c|c|}  \hline
$l$	 & $B_{0} \times 10^{19}$ & $B_{1}\times 10^{19}$  & $B_{2}\times 10^{19}$ & $B_{3}\times 10^{19}$ & $B_{4}\times 10^{19}$ & $B_{5}\times 10^{19}$ & $B_{6}\times 10^{19}$ & $B_{7}\times 10^{19}$ & $B_{8}\times 10^{19}$  \\  \hline
 $e$ & 1.10 & 0.04 & 0.006 & -0.05 & 0.006 & 0.20 & 0.03 & -0.14 & 0.03  \\
  $\mu$ & 1.09 & 0.04 & 0.006 & -0.05 & 0.006 & 0.20 &0.03 & -0.14 & 0.03 \\
  $\tau$ & 0.40 & 0.030 & 0.004 & -0.008 & 0.001 & 0.15 & 0.02 & -0.025 & 0.005\\
 \hline
\end{tabular}
\caption{The values of  $B^{l}_{i}$'s for different lepton flavors.}
\label{Bvalues}
\end{center}
\end{table*}
}

These factors are calculated using the form-factors given in ref. \cite{Gubernari:2018wyi,Bailey:2015dka} which is also used in the  {\tt flavio} package. The different values of functions $A_{i}^{l}$'s and $B_{i}^{l}$'s are due to the differences in the masses of the leptons. These functions determine the values of the flavor ratios. For different scenarios, depending on the presence of a type(s) of new physics WCs, the behaviour of the ratio $R^{\tau \mu}$ is different as discussed below: 
\begin{itemize}
 \item{{\bf{SU-I:}} In this scenario,  only $C_{9}^{U}$ is present. The ratio $R_{K}^{\tau \mu}$ takes the following form,
\begin{equation}
R_K^{\tau\mu}=\frac{A_{0}^{\tau}+  A_{1}^{\tau}\, C_{9}^{U}  + A_{2}^{\tau}\,(C_{9}^{U} )^{2}}{A_{0}^{\mu}+  A_{1}^{\mu}  \,C_{9}^{U}  + A_{2}^{\mu}\, (C_{9}^{U} )^{2}} \,.
\end{equation}
 As can be seen from eq. \ref{A}, the $A_i$'s functions take different set of values for $\mu$ and $\tau$.  As $ A_{1}^{\mu}> A_{1}^{\tau}$ and $ A_{2}^{\mu}> A_{2}^{\tau}$, $R_K^{\tau\mu}$ deviates from its SM value even for universal couplings. For 
 the negative values of $C_{9}^{U}$, the denominator becomes smaller as compared to the numerator resulting in larger values of $R_{K}^{\tau \mu}$ as compared to the SM. As the value of  $C_{9}^{U}$ increases, the difference between the numerator and denominator stars decreasing. The difference becomes almost negligible for $C_{9}^{U} \approx 2.4$, yielding $R_{K}^{\tau \mu} \approx 1$ which is less than its SM value.
The value of the ratio $R_{K}$ relinquishes its SM value which is $\approx 1$, even for the sufficiently larger values of $C_{9}^{U}$. This is due to the fact that  the functions $ A_{1}^{\mu}$ and $ A_{2}^{\mu}$ are nearly equal to their electron counterparts i.e., $A_{1}^{e}$ and $ A_{2}^{e}$.}

On the other hand, the ratio $R_{K^*}^{\tau\mu}$ can be written as 
\begin{equation}
R_{K^*}^{\tau\mu}=\frac{B_{0}^{\tau}+  (B_{1}^{\tau}+B_{5}^{\tau})  C_{9}^{U}  + (B_{2}^{\tau}+B_{6}^{\tau})(C_{9}^{U} )^{2}}{B_{0}^{\mu}+  (B_{1}^{\mu}+B_{5}^{\mu})  C_{9}^{U}  + (B_{2}^{\mu}+B_{6}^{\mu})(C_{9}^{U} )^{2}}.\\
\end{equation}
It is evident from  table \ref{Bvalues} that the prefactors $B^{l}_{i}$'s are not identical for $\tau$ and $\mu$ as they are in the case of $e$ and $\mu$ resulting in the different behaviour  of $R_{K^*}^{\tau\mu}$ as compared to the SM like behaviour of  $R_{K^*}$, even for the universal NP couplings. As $C_{9}^{U}$ increases in the positive side, the rate of increment of the  linear and quadratic terms of the numerator with respect to its constant term, is larger than that of the denominator. This is because of the fact that the constant term of denominator is already larger than the linear and quadratic terms by one and two order of magnitudes, respectively. On the other hand as $C_{9}^{U}$ goes in the negative direction, the same argument is valid, reducing the value of $R_{K^*}^{\tau\mu}$. 

\item{{\bf{SU-II:}} In this scenario both $C_{9}^{U}$ and $C_{10}^{U}$ are present having the correlation $C_{9}^{U}=-C_{10}^{U}$. Now the ratio $R_{K}^{\tau\mu}$ can be written as, 
\begin{equation}
R_K^{\tau\mu}=\frac{A_{0}^{\tau}+  (A_{1}^{\tau}-A_{3}^{\tau})C_{9}^{U}  + (A_{2}^{\tau}+A_{4}^{\tau})(C_{9}^{U} )^{2}}{A_{0}^{\mu}+ (A_{1}^{\mu}-A_{3}^{\mu})  C_{9}^{U}  + (A_{2}^{\mu}+A_{4}^{\mu}) (C_{9}^{U} )^{2}}.
\end{equation}
Here all $A_{i}$ functions corresponding to $\tau$ and $\mu$ contribute to the ratio $R_K^{\tau\mu}$. The observable $R_K^{\tau\mu}$ does not deviate much from its SM value for universal  new physics couplings, as  $ (A_{2}^{\tau}+A_{4}^{\tau}) \approx  (A_{2}^{\mu}+A_{4}^{\mu})$ and  $(A_{1}^{\tau}-A_{3}^{\tau})$ \& $(A_{1}^{\mu}-A_{3}^{\mu})$ are only marginally different from each other. Again the similar nature of the functions for $\mu$ and $e$ owing to their masses, make the ratio $R_K$ to surrender to its SM value.}

Similarly, for the $B\to K^{*}\ell^{+}\ell^{-}$ decay mode, the flavor ratio takes the following form,
\begin{equation}
R_{K^*}^{\tau\mu}=\frac{B_{0}^{\tau}+  \xi^{\tau}_{L} C_{9}^{U}  + \xi^{\tau}_{Q}(C_{9}^{U} )^{2}}{B_{0}^{\mu}+ \xi^{\mu}_{L} C_{9}^{U}  + \xi^{\mu}_{Q} (C_{9}^{U} )^{2}},
\end{equation}
where $\xi^{l}_{L} = (B_{1}^{l}+B_{5}^{l}-B_{3}^{\l}-B_{7}^{l})$ and $\xi^{l}_{Q} = (B_{2}^{l}+B_{4}^{l}+B_{6}^{l}+B_{8}^{l})$.  In this scenario, for positive values of $C_{9}^{U}$, both the numerator and denominator show a similar rate of increment in the sum of their linear and quadratic terms. This leads to the ratio being nearly constant in the positive region, aligning with the SM value. However, as the values of $C_{9}^{U}$ become increasingly negative, the muonic quadratic term ($\xi^{\mu}_{Q}$) compensates for the reduction in the linear term ($\xi^{\mu}_{L}$) more effectively than its tau counterpart. Consequently, this results in a marginal decrease in the value of $R_{K^*}^{\tau\mu}$ for $C_{9}^{U} <0$.

\item{{\bf{SU-III:}} This scenario consists of $ C_{9}^{U}$ and its right handed counterparts $ C_{9}^{'U}$ with the correlation  $C_{9}^{U}=-C_{9}^{'U}$. It is evident from eq. \ref{gamma} that for $R_K^{\tau\mu}$, the contribution arising from the new physics effects vanishes rendering only the SM contribution. The same  is true for $R_K$ as well making it inert to $C_{9}^{U}=-C_{9}^{'U}$ new physics effects.}

However, in contrast to $R_{K}^{\tau\mu}$, the effects of new physics do not tend to disappear in the ratio $R_{K^*}^{\tau\mu}$. This distinction is evident in equation \ref{gamma1}, where terms linear and quadratic in $(C_{9}^{U} - C_{9}^{'U})$ are present, unlike equation \ref{gamma}. For this scenario, the ratio $R_{K^*}^{\tau\mu}$ takes the form,
\begin{equation}
    R_{K^*}^{\tau\mu} = \frac{B_{0}^{\tau}+2B_{5}^{\tau} C_{9}^{U}+4B_{6}^{\tau}(C_{9}^{U} )^{2}}{B_{0}^{\mu}+2B_{5}^{\mu} C_{9}^{U}+4B_{6}^{\mu}(C_{9}^{U} )^{2}}\,.
\end{equation}
For the positive as well as negative values of $ C_{9}^{U}$, the behaviour of  $R_{K^*}^{\tau\mu}$ can be understood with the same reasoning as for the SU-I scenario. 

\end{itemize}

From Fig.~\ref{fig:tau-mu-1}, it is therefore evident that unlike $R_{K}$ and $R_{K^*}$,  the LFU ratios $R_{K}^{\tau \mu}$ and $R_{K^*}^{\tau \mu}$ may render values different from their SM predictions even for class of new physics solutions having only universal component. Therefore mere deviation of these observables from the SM cannot confirm the nature of new physics in $\tau-\mu$ sector, i.e if any experiment measures $R_{K}^{\tau \mu}$ and $R_{K^*}^{\tau \mu}$ with a value different from their SM predictions, we cannot jump into the conclusion that this deviation is due to LFUV type of new physics. For such a discrimination, additional analysis would be required. A simple method would be to obtain the extremum values of $R_{K}^{\tau \mu}$ and $R_{K^*}^{\tau \mu}$ for the class of solutions obtained under the assumption of only universal couplings and compare this with the allowed range obtained for solutions having both the components. If the two regions are distinct, $R_{K}^{\tau \mu}$ and $R_{K^*}^{\tau \mu}$ can enable discriminating between the universal and non-universal type of new physics. In the following we discuss this possibility for the new physics scenarios allowed by the current experimental data in $b \to s \ell \ell$ ($\ell=e,\, \mu $) sectors.

The 1$\sigma$ predicted range of  $R_{K}^{\tau \mu}$ and $R_{K^*}^{\tau \mu}$ for all allowed solutions in F-I and F-II frameworks are depicted in Fig.~\ref{fig:tau-mu-2}. It is evident from the top panel of the figure that none of the F-I solutions can provide large enhancement of $R_{K}^{\tau \mu}$ observable above the SM value. The SU-I, i.e $C_9^U<0$ solution can only provide a marginal enhancement ($\lesssim 5\%$) whereas the other two F-I solutions relinquish $R_{K}^{\tau \mu}$ within its SM range. On the other hand, almost all F-II solutions can enhance $R_{K}^{\tau \mu}$ above the SM prediction. The enhancement can be large, up to $\sim 25\%$  for the S-V solution. This implies that the observation  of $R_{K}^{\tau \mu}$ with value $\gtrsim 10\%$ above the SM prediction  would be possible only for the class of solutions having both universal as well as non-universal couplings.  For e.g., the measurement of $R_{K}^{\tau \mu}$ with a value $\gtrsim 1.5$ with an absolute uncertainty of 0.1 can lead to a 2$\sigma$ distinction between the two classes of solutions.

The  $R_{K^*}^{\tau \mu}$  predictions  for F-II solutions were first obtained in \cite{SinghChundawat:2022zdf}. These along with the predictions for F-I solutions are demystified in the bottom panel of Fig.~\ref{fig:tau-mu-2}. It is perceptible from the figure that that for all F-I solutions, the predicted values of $R_{K^*}^{\tau \mu}$ is less than the SM value whereas a number of solutions in F-II framework predict $R_{K^*}^{\tau \mu}$ greater than the SM. In particular the S-V and S-XIII can lead to a large enhancement in $R_{K^*}^{\tau \mu}$ over the SM.  Therefore if $R_{K^*}^{\tau \mu}$ is measured with a value greater than the SM, this will not only confirm the presence of new physics but will also reveal its non-universal nature. For e.g., if $R_{K^*}^{\tau \mu}$ is measured  with a value $\sim$ 20\% above the SM prediction ($\approx$ 0.40) with an absolute uncertainty of 0.04, the new physics solutions with non-universal component will be favoured over new physics with universal solutions at the level of 2$\sigma$. However, if the measured value is less than the SM prediction, it would be difficult to reveal the nature of new physics through $R_{K^*}^{\tau \mu}$.

Currently the study of $b \to s \tau^+ \tau^-$ decays are restricted  from the experimental side due to intricacy in  reconstruction of tauons in the final states. Because of this, at present, we only have upper bounds in this sector which are several orders of magnitude above the SM predictions. For instance, the upper bounds on the branching ratios of $B \to K \tau^+ \tau^-$ and $B \to K^* \tau^+ \tau^-$ decays are 2.25 $\times 10^{-3}$ \cite{BaBar:2016wgb} and $2 \times 10^{-3}$ \cite{Belle:2021ndr}, respectively. Therefore in order to utilize the potential of these decays \cite{Alonso:2015sja,Belle-II:2018jsg} (as well as $b \to d \tau^+ \tau^-$ decays \cite{Ali:2023kvz}) for investigating new physics, a drastic improvement in tau-reconstruction techniques will be required. We hope that such sensitivities would be achieved at the  HL-LHC \cite{LHCb:2018roe}, Belle II \cite{Belle-II:2018jsg} and FCC-ee experiments \cite{Bernardi:2022hny,Kamenik:2017ghi,Li:2020bvr}. Based on the current sensitivity analysis, the HL-LHC and Belle-II can detect $B \to K \tau^+ \tau^-$ and  $B \to K^* \tau^+ \tau^-$ decays up to a level of $\sim(10^{-4} - 10^{-5})$ whereas owing to the high precision vertex reconstruction, the FCC-ee experiment can not only perform  a precision measurement of the branching ratio of $B \to K^* \tau^+ \tau^-$ decay up to the SM level but can also  measure its angular distribution.

%%%%%%%%%%%%%%%%%
 \section{Conclusions}
\label{concl}
%%%%%%%%%%%%%%%%%%%%%%%%%%%
The ratios $R_{K}^{\tau \mu} $ and $ R_{K^{*}}^{\tau \mu} $ are expected to provide  tests of LFU violation in the  $\tau-\mu$ sector in the same way  as $R_{K} $ and $ R_{K^{*}} $ observables furnish in the $\mu-e$ sector. However, we find that these ratios 
relinquish values different from their SM predictions even for new physics with only universal couplings. A good fit to the current  $b \to s \ell \ell$ ($\ell=e,\,\mu$) data can be provided by assuming new physics couplings to be universal in nature as well as for the scenarios having both universal as well as non-universal components. Therefore bare deviation of $R_{K}^{\tau \mu} $ and $ R_{K^{*}}^{\tau \mu} $ from their SM values cannot confirm the nature of possible new physics.  A careful anatomization of these observables for the solutions corresponding to the two classes of new physics will be required to identify the  new physics type.  By comparing the predictions of $R_{K^{(*)}}^{\tau \mu}$ for the current allowed solutions,   we find  that two classes of solutions can be discriminated if the measured value of  $ R_{K^{*}}^{\tau \mu} $ is greater than the SM prediction. 

\bigskip
\noindent
{\bf Acknowledgements}: The work of A.K.A. is supported by SERB-India Grant CRG/2020/004576.

\end{document}